\documentclass[final,english]{bullsrsl}[2022/06/15]

\usepackage[latin1]{inputenc}
\usepackage[T1]{fontenc}

\usepackage{natbib} 
\usepackage{graphicx}

\begin{document}
\title{Surface Brightness Properties of LSB Galaxies with the International Liquid Mirror Telescope}

\author[affil={1}, corresponding]{Jiuyang}{Fu}
\author[affil={2,3}]{Bhavya}{Ailawadhi}
\author[affil={4,5}]{Talat}{Akhunov}
\author[affil={6}]{Ermanno}{Borra}
\author[affil={2,7}]{Monalisa}{Dubey}
\author[affil={2,7}]{Naveen}{Dukiya}
\author[affil={1}]{Baldeep}{Grewal}
\author[affil={1}]{Paul}{Hickson}
\author[affil={2}]{Brajesh}{Kumar}
\author[affil={2}]{Kuntal}{Misra}
\author[affil={2,3}]{Vibhore}{Negi}
\author[affil={2,8}]{Kumar}{Pranshu}
\author[affil={1}]{Ethen}{Sun}
\author[affil={9}]{Jean}{Surdej}

\affiliation[1]{Department of Physics and Astronomy, University of British Columbia, 6224 Agricultural Road, Vancouver, BC V6T 1Z1, Canada}
\affiliation[2]{Aryabhatta Research Institute of Observational sciencES (ARIES), Manora Peak, Nainital, 263001, India}
\affiliation[3]{Department of Physics, Deen Dayal Upadhyaya Gorakhpur University, Gorakhpur, 273009, India}
\affiliation[4]{National University of Uzbekistan, Department of Astronomy and Astrophysics, 100174 Tashkent, Uzbekistan}
\affiliation[5]{ Ulugh Beg Astronomical Institute of the Uzbek Academy of Sciences, Astronomicheskaya 33, 100052 Tashkent, Uzbekistan}
\affiliation[6]{Department of Physics, Universit\'{e} Laval, 2325, rue de l'Universit\'{e}, Qu\'{e}bec, G1V 0A6, Canada}
\affiliation[7]{Department of Applied Physics, Mahatma Jyotiba Phule Rohilkhand University, Bareilly, 243006, India}
\affiliation[8]{Department of Applied Optics and Photonics, University of Calcutta, Kolkata, 700106, India}
\affiliation[9]{Institute of Astrophysics and Geophysics, University of Li\`{e}ge, All\'{e}e du 6 Ao$\hat{\rm u}$t 19c, 4000 Li\`{e}ge, Belgium}

\correspondance{fujyang2@student.ubc.ca}
\date{12th May 2023}
\maketitle

% \author[affil1]{FirstName (+ MiddleInitials if necessary)}{FamilyName}
% \author[affil2]{...}{}
% \equalcontribauthor[]{}{} % Maximum two --> counter
% \consortium[affil]{Consortium Name}
% With consortium: affiliation will be set to "See Appendix 1 for a full
% list of consortium members and their respective affiliations
% \affiliation[affil1]{...}
% \affiliationq[affil2]{...}

% \correspondence[]{}
% No explicit corresponding author: use first author
% 

% Abstract of the paper in the same language as the paper
\begin{abstract}
Low surface brightness (LSB) galaxies make up a significant fraction of the luminosity density of the local universe. Their low surface brightness suggests a different formation and evolution process compared to more-typical high-surface-brightness galaxies. This study presents an analysis of LSB galaxies found in images obtained by the International Liquid Mirror Telescope during the observation period from October 24 to November 1, 2022. 3,092 LSB galaxies were measured and separated into blue and red LSB categories based on their  $g'-i'$ colours. In these samples, the median effective radius is 4.7 arcsec, and the median value of the mean surface brightness within the effective radius is 26.1 mag arcsec$^{-2}$. The blue LSB galaxies are slightly brighter than the red LSB galaxies. No significant difference of ellipticity was found between the blue and the red LSB galaxies.
\end{abstract}

\keywords{LSB galaxy, surface brightness}

%%%%%%%%%%%%%%%%%%%%%%%%%%%%%%%%%%%%%%%%%%%%%%%%%%
\section{Introduction}

Galaxies can be classified into two broad groups based on their colors in optical light, which are strongly linked to the types of stars that dominate their populations. This classification is closely related to the overall structure, or morphology, of the galaxies, resulting in a separation into ``red'' and ``blue'' sequences \citep{Blanton_2009}. Although extensive surveys like SDSS have provided opportunities to examine the colour patterns of galaxies in great detail \citep{Balogh_2004}, our understanding of colour distribution at lower surface brightness levels remains limited.

Currently deeper surveys have just been taken place to produce large LSB galaxies catalogs, one of which is the International Liquid Mirror Telescope (ILMT) survey that will bridge the gap in our understanding of the LSB galaxies population. The ILMT is a zenith-pointing telescope installed at Devasthal Peak in the central Himalayan range in India. Since the ILMT cannot track stellar objects like conventional glass mirror telescopes, tracking is accomplished electronically. This is achieved by using time-delay integration (TDI) readout, in which every pixel in the image results from an integration over 4096 pixels along each column of the CCD. This reduces flat-field variations by a factor of $\sim$ 4.5 magnitudes \citep{Kumar_2018}.

LSB galaxies are typically difficult to detect due to their low intensity. With a $22\times22$-arcmin$^{2}$ field of view, the ILMT can survey $\sim 50$ square degrees of sky each night. Another advantage of the ILMT is that it always observes at the zenith, so the sky brightness and atmospheric absorption and scattering are minimized. Images from different nights can be co-added to increase the signal-to-noise ratio. The ILMT employs $g'$, $r'$ and $i'$ filters, which makes it possible to study the difference between red and blue LSB galaxies based on $g'-i'$ colour.  

\section{Observations and analysis}

Six images of the same target area captured by the ILMT during the observation period from October 24 to November 1, 2022, were analyzed. Each of the $g'$, $r'$, and $i'$ filters was employed for two nights, respectively. The size of each image is 4096 pixels $\times$ 36817 pixels. Given that the pixel size of the CCD camera is about 0.328 arcsec, the scanned area in each image is 22.3 arcmin $\times$ 201.1 arcmin. The coordinate ranges of the area studied are 07:14:14.03$-$07:27:38.31 in right ascension and 29:12:43.3$-$29:34:59.5 in declination (J2000).

\begin{figure}[t]
\centering
%\begin{tabular}[b]{c}
%\includegraphics[width=.30\linewidth]{original} \\
%(a) Original Image
%\end{tabular} 
%\begin{tabular}[b]{c}
%\includegraphics[width=.30\linewidth]{segmented} \\
%(b) Cleaned Image
%\end{tabular} 
%\begin{tabular}[b]{c}
%\includegraphics[width=.30\linewidth]{original-segmented} \\
%(c) Subtracted Image 
%\end{tabular}
\includegraphics[width=\textwidth]{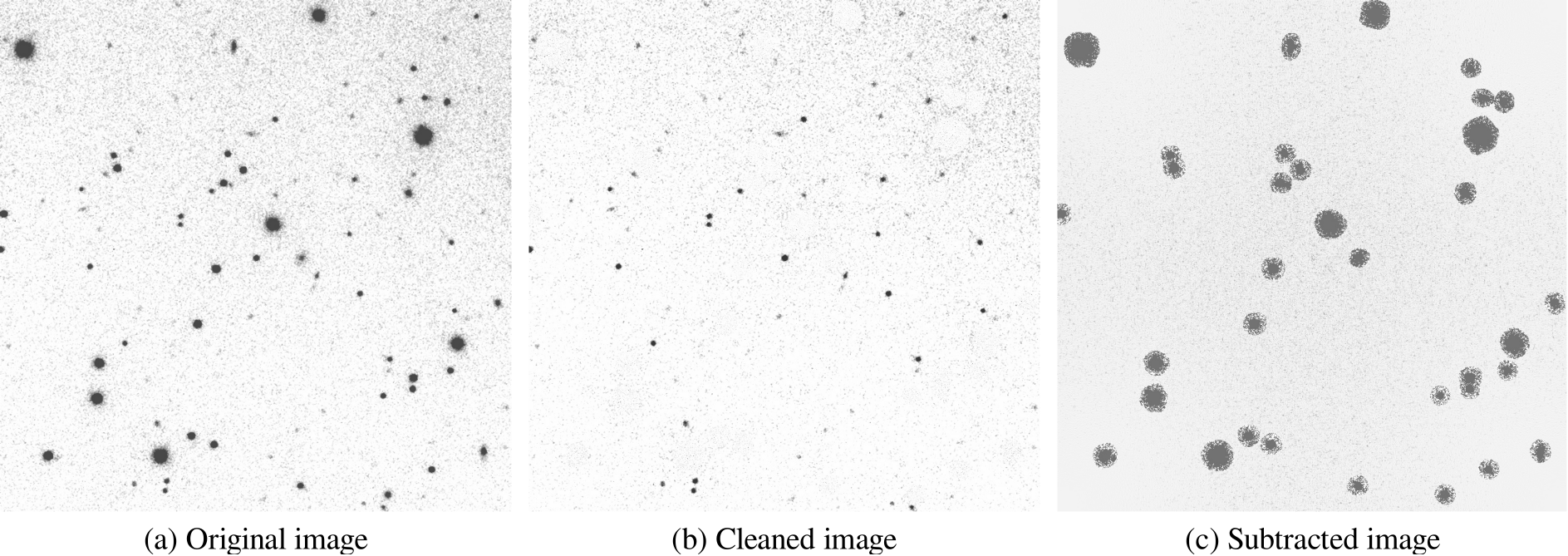}
\smallskip

\begin{minipage}{12cm}
\caption{The original image from the ILMT is shown in panel (a). In panel (b), the bright sources that are 4 standard deviation above the median background noise were cleaned if their size was larger than 1257 arcsec$^{2}$. The subtracted bright sources are shown in panel (c).}
\label{fig:example}
\end{minipage}
\end{figure}

Faint objects such as LSB galaxies may be obscured by brighter sources in the image. This can make it difficult to detect and analyze the faint objects, and may even result in false detections or misinterpretation of the data. In this work, bright sources and their associated diffuse light were cleaned using Astropy's Photutils detect$\_$sources program augmented by transformations using binary masks. All sources that are 4 standard deviations above the median of the sky background and have an area greater than 1257 arcsec$^{2}$ (corresponding to a circular area with radius of 20 arcsec) were replaced with the median background. For each bright source, the radius of the region to be replaced is the radius at which the median intensity drops to the level of the median background noise. With this method, all diffuse light of a bright source can be removed, which makes it more accurate during the process of local background estimation. An example of the imaging cleaning process is shown in Fig.\,\ref{fig:example}.

We classified a galaxy as LSB if its $r'$-band disc central surface brightness is fainter than 21 mag arcsec$^{-2}$. This particular threshold was selected because the brightness of the dark sky at the location of the ILMT is 21.09 mag/arcsec$^{2}$ in the $r'$-band, given the definition that LSB galaxy is fainter than the dark sky. While the circular aperture may be effective for objects with limited angular size or spherical shape, it is not the most suitable option for galaxies that tend to have extensive angular size, unconventional structure, or an edge-on configuration \citep{Du_2015}. Therefore, the surface brightness modelling of our LSB galaxy samples was done by SExtractor software using elliptical apertures. The Sersic model was used to fit the intensity profiles of our LSB galaxies. The effective radius $r_{\rm eff}$ is measured along the major axis. The surface brightness $\mu_{\rm eff}$ is measured within the effective radius, which has an elliptical shape with the same ellipticity as that of the aperture.

\section{Results}

\begin{figure}[t]
\centering
\includegraphics[width=0.7\textwidth]{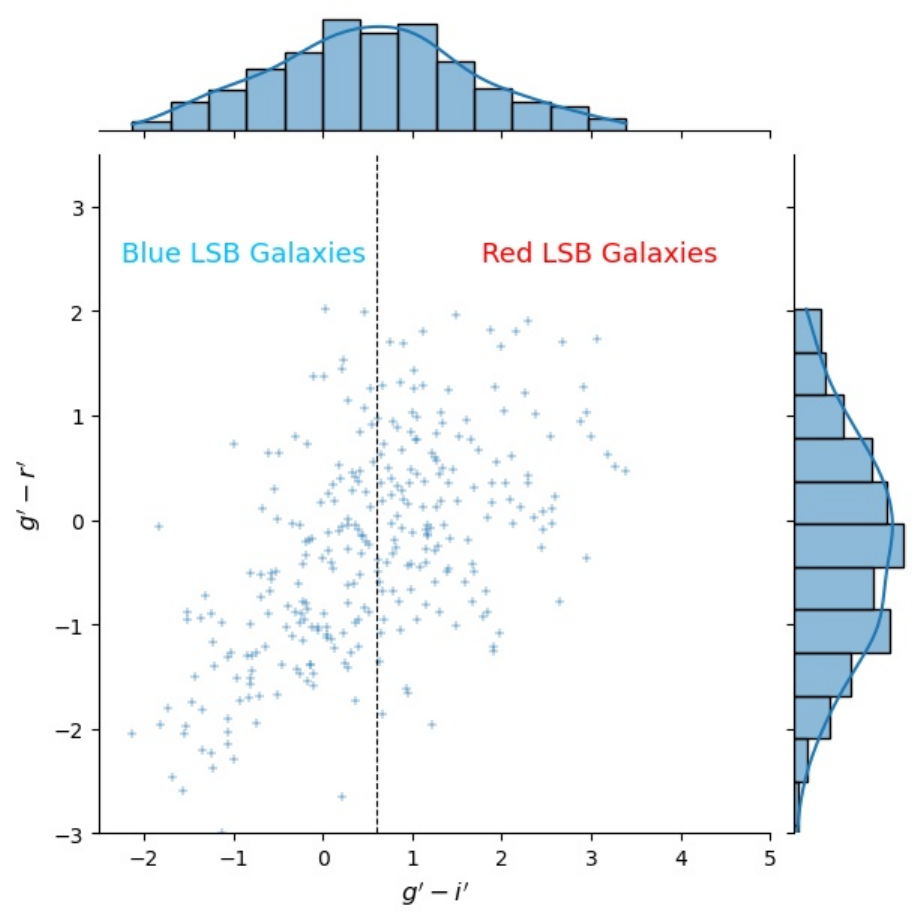}

\begin{minipage}{12cm}
\caption{Colour-colour diagram of LSB galaxy samples. The galaxies were divided into red ($g'-i' \geq 0.66$) and blue ($g'-i' < 0.66$) subsamples. The dividing colour is determined by the median value of $g'-i'$ \citep{Greco_2018}. The distributions of $g'-i'$ and $g'-r'$ are shown in the histograms on the marginal axes. The blue lines overlaid on the histograms show the density distributions fit by kernel density estimation.}
\label{fig:ILMT}       % Give a unique label to the figure. 
\end{minipage}
\end{figure}

A total of 3092 potential LSB galaxies were identified and measured. The coverage of the central surface brightness in $r'$ band of all the LSB galaxy samples is from 21.0 to 26.1 mag arcsec$^{-2}$. 299 of these are shown in the colour-colour diagram of Fig.\,\ref{fig:ILMT}. It was found that the $g'-r'$ and $g'-i'$ are positively correlated. The distributions of $g'-i'$ and $g'-r'$ are shown in the histograms on the marginal axes with the fitted kernel density estimation (KDE) curves displayed as the blue curves. These 299 LSB galaxies were separated into blue and red LSB categories. The dividing colour of 0.66 corresponds to the median value of $g'-i'$ (displayed as the black dashed line in Fig.\,\ref{fig:ILMT}). A comparison between the two categories is given in Table\,\ref{table:1}. The blue LSB galaxies are slightly brighter than the red LSB galaxies. No significant difference of ellipticity was found between the two LSB categories.

\begin{table}[h!]
\centering
\caption{Statistical Results of LSB Galaxy Samples in $r'$ band}
\bigskip

\begin{tabular}{lcc} 
\hline
\textbf{Measurement} & \textbf{Blue LSB Galaxy} & \textbf{Red LSB Galaxy} \\
\hline
Detected Number                         & 150              & 149              \\
Median $r_{\rm eff}$ (arcsec)             &  4.62 $\pm$ 0.76 &  4.74 $\pm$ 0.77 \\
Median $\mu_{\rm eff}$ (mag arcsec$^{-2}$) & 25.93 $\pm$ 0.38 & 26.34 $\pm$ 0.38 \\
Mean Ellipticity                        &  0.16 $\pm$ 0.03 &  0.18 $\pm$ 0.03 \\ 
\hline
\end{tabular}
\label{table:1}
\end{table}

\section{Discussion}
 
The light profiles for red LSB galaxies are typically smooth, and they can be well-described by the Sersic function, while blue LSB galaxies tend to have irregular morphologies. Only Sersic model was used in fitting light profiles. Improvements by using a more complete model are considered in further research.

The colour-colour diagram has a higher dispersion than expected. The main reason for this is the low confidence of the star-galaxy separation. When more images of the same target area are obtained, they will be co-added to improve the star-galaxy separation. The evolution path of LSB galaxies, and the age and metallicity could then be studied.

While 3092 LSB galaxies were detected in the $r'$ band, less than 10$\%$ of them were detected in the $g'$, $r'$ and $i'$ bands simultaneously. One reason for this is that some sample LSB galaxies are too faint to be detected in the $g'$ and $i'$ bands. This is reasonable because either in the colour-colour diagram that obtained here, or in published colour-colour diagrams, more LSB galaxies have a positive $g-i$ value and are thus fainter in the $g'$ band. It is also possible that some detected LSB galaxies are not real celestial objects, which may either be cosmic ray hits or local maxima occurred in the background noise. 

\begin{acknowledgments}
The 4m International Liquid Mirror Telescope (ILMT) project results from a collaboration between the Institute of Astrophysics and Geophysics (University of Li\'{e}ge, Belgium), the Universities of British Columbia, Laval, Montreal, Toronto, Victoria and York University, and Aryabhatta Research Institute of observational sciencES (ARIES, India). The authors thank Hitesh Kumar, Himanshu Rawat, Khushal Singh and other observing staff for their assistance at the 4m ILMT.  The team acknowledges the contributions of ARIES's past and present scientific, engineering and administrative members in the realisation of the ILMT project. JS wishes to thank Service Public Wallonie, F.R.S.-FNRS (Belgium) and the University of Li\'{e}ge, Belgium for funding the construction of the ILMT. PH acknowledges financial support from the Natural Sciences and Engineering Research Council of Canada, RGPIN-2019-04369. PH and JS thank ARIES for hospitality during their visits to Devasthal. B.A. acknowledges the Council of Scientific $\&$ Industrial Research (CSIR) fellowship award (09/948(0005)/2020-EMR-I) for this work. M.D. acknowledges Innovation in Science Pursuit for Inspired Research (INSPIRE) fellowship award (DST/INSPIRE Fellowship/2020/IF200251) for this work.
\end{acknowledgments}

\begin{furtherinformation}

%\begin{orcids}
%\orcid{0000-1111-2222-3333}{Hàrry}{Harrisòn}
%\orcid{1111-2222-3333-4444}{Leonie}{van Leon}
%\orcid{2222-3333-4444-5555}{Lotta}{Lothardis}

%{\sl This section is optional.
%You may list here the ORCIDs of those authors who would like to share them, one per line, with the \verb|\orcid{|\texttt{\emph{ORCID}}\verb|}{|\texttt{\emph{First name}}\verb|}{|\texttt{\emph{Last name}}\verb|}| command.
%This command typesets the information, and makes the ORCIDs themselves active links to the corresponding records on \href{https://orcid.org}{orcid.org}.

%Unlike in this sample, no other text should actually be included here and this section should reduce to a bare list.
%The \verb|\orcid| command controls line feeds by itself; please do not insert any \verb|\\| or \verb|\newline| before or after them.}
%\end{orcids}

\begin{authorcontributions}
%This section is mandatory when there is more than one author.
%The contributions of each author (identified by their initials) must be declared.
%We recommend to follow the \href{http://credit.niso.org}{CRediT} taxonomy (Contributor Roles Taxonomy).
This work results from a long-term collaboration to which all authors have made significant contributions.
\end{authorcontributions}

\begin{conflictsofinterest}
%This section is \emph{mandatory}.
%Authors must declare any personal or professional circumstances that may be perceived as influencing the research reported in the paper.
%If there is no conflict of interest, please state that 
The authors declare no conflict of interest.
\end{conflictsofinterest}

\end{furtherinformation}

\bibliographystyle{bullsrsl-en}

\bibliography{S11-P15_FuJ}

\end{document}